\documentclass{iaus}
\usepackage{graphicx}
\usepackage{natbib}

\title[Type II bursts, CMEs \& Flare parameters] %% give here short title %%
{Relation Between Coronal Type II Bursts, Associated Flares and CMEs \\  }

\author[George Pothitakis \& al]   %% give here short author list %%
{George Pothitakis$^1$, Panagiota Preka-Papadema$^1$, Xenophon Moussas$^1$, Constantine Caroubalos$^2$,
Constantine Alissandrakis$^4$, Panagiotis Tsitsipis$^3$, Athanasios Kontogeorgos$^3$ \and Alexander Hillaris$^1$}

\affiliation{$^1$Department of Physics, University of Athens, 15784 Athens, Greece\\[\affilskip]
$^2$Department of Informatics, University of Athens, 15783 Athens, Greece\\[\affilskip]
$^3$Department of Electronics, Technological Education Institute of Lamia, Lamia, Greece\\[\affilskip]
$^4$Department of Physics, University of Ioannina, 45110 Ioannina, Greece\\[\affilskip]
}

\pubyear{2008}
\volume{257}  %% insert here IAU Symposium No.
\jname{Universal Heliophysical Processes}
\editors{A.C. Editor, B.D. Editor \& C.E. Editor, eds.}
\begin{document}

\maketitle

\begin{abstract}
We study a sample of complex events; each includes a coronal type II burst, accompanied by a 
GOES SXR flare and LASCO CME. The radio bursts were recorded by the ARTEMIS-IV radio 
spectrograph (100-650 MHz range); the GOES SXR flares and SOHO/LASCO CMEs, 
were obtained from the Solar Geophysical Data (SGD)
and the LASCO lists respectively. The radio burst-flare-CME characteristics were compared and
two groups of events with similar behavior were isolated. 
In the first the type II shock exciter appears to be a flare blast wave
propagating in the wake of a CME. In the second the type II burst appears CME initiated
though it is not always clear if it is driven by the bow or the flanks of the CME or
if it is a reconnection shock.

\end{abstract}

\firstsection % if your document starts with a section, remove some space above using this command.
\section{Introduction}

The MHD shock radio signatures, in the interplanetary medium and the solar corona,
are the kilometric type II bursts and the  metric type II radio emissions respectively.
Though the former have been, unambiguously, identified with shocks piston-driven by CMEs, the exciter 
of the latter is somewhat more controversial as it can be either a blast-wave or 
a CME driven shock. Only the CME driven shocks are expected to
propagate into interplanetary space; the blast-waves are damped with distance and,
probably, rarely escape the lower corona (cf. \citet{Gopalswamy06} also \citet{Pick06} and references therein).
This ambiguity about the exciter of coronal type II bursts has initiated a number of publications 
(cf. \citet{Kahler1984}, \citet{Classen}, etc.) in which the CME, flare and type II parameters are compared. 
In this report we examine a set of complex events in search of groups with similarities as regards the relationship of
type II burst-flare-CME characteristics; each is expected to represent a different shock generation processes.

\section{Data Selection and Analysis}

The ARTEMIS IV radiospectrograph (\cite{Caroubalos01})
observed  40 type II and/or IV radio bursts (1998-2000) which were published in the form of a
catalogue \citep{Caroubalos04}. The gross spectral characteristics of these events, and the associated CME and flare
parameters were summarized in this catalogue; here we adopted the same numbering of events.

From the original catalogue we selected fourteen events; (Table 1); they all include flares 
associated with a type II ARTEMIS-IV metric burst and a SOHO/LASCO CME. 

%-------------------------------------------------------------------------------------------------------------------
\begin{table}
  \begin{center}
  \caption{Characteristic Parameters of Each Event in the Data Set}
  \label{tab1}
 {\scriptsize
  \begin{tabular}{|l|c|c|c|c|c|}
\hline
{\bf Event} 	& $\delta\tau$ (min)& $V_r$		& $D$ (min)		& $V_{II}$ (km/sec) &$\Delta T$ (min)\\
\hline
{\bf Group I}	& 			& 			& 			& 			&		\\
\hline
19		      &  -22		& 3.12		& 3		      & 1213		& -4		\\
24		      &  -12		& 2.50		& 3		      & 1940		& -5		\\
25		      &  -32		& 1.77		& 1		      & 1477		& -4		\\
\hline
{\bf Average}	& -22$\pm$5.2	& 2.5 $\pm$0.4  	& 2.33$\pm$0.67	& 1543$\pm$212	& -4.3$\pm$0.33	\\	
\hline
{\bf Group II}		& 		& 			& 			& 		&		\\ 
\hline
08			& 2		& 0.38			& 11			& 416		& -1	   \\
21			& -10		& 0.95			& 3			& 430		& -1	   \\
23			& 20		& 0.80			& 11			& 806		& -1	   \\
27			& 9		& 1.07 			& 1			& 375		& -1	   \\
30			& 17		& 1.73			& 3			& 737		& 0	\\
32			& 17		& 0.82			& 4			& 442		& 0	   \\
39			& 2		& 0.65			& 7			& 494		& -1	   \\
40			& 19		& 1.12			& 6			& 598		& -2	 \\ 
\hline
{\bf Average}	& 9.5$\pm$3.8	& 0.9$\pm$0.14  	& 5.8$\pm$1.3	& 537$\pm$57	& 0.9$\pm$0.2	\\	
\hline
{\bf Unclassified}& 			& 			&			&			&	\\ 
\hline
06			& -13		& 1.00 			& 10			& 940			& -5	 \\ 
33			& -4		& 1.02			& 5			& 1300		& -3	 \\
36			& 22		& 0.85			& 17			& 1430		& -3	 \\
\hline
{\bf Overall Average}		& 1.1$\pm$4.6	& 1.27 $\pm$0.20  	& 6.1$\pm$1.24 & 900$\pm$132	& -2.2$\pm$0.5	\\	 \hline
  \end{tabular}
  }
 \end{center}
\end{table}
%-------------------------------------------------------------------------------------------------------------------

The parameters used in our study were:

\begin{itemize}
\item {The Type II speed, $V_{II}$, in km/sec. They were calculated from the frequency drift rate of the
type II bands, assuming a Newkirk \citep{Newkirk} corona and radial propagation of the MHD shock.}

\item {The ratio of the Type II speed to the CME speed, $V_{r} =V_{II}/V_{CME}$).
$V_{CME}$ was obtained from the on line LASCO lists}

\item {The time interval, $\delta \tau$, between the CME liftoff and the flare onset
from the extrapolated value cited in the LASCO lists and 
the GOES SXR profiles respectively.}

\item {The Type II duration in minutes, $D$, from the ARTEMIS-IV dynamic spectra.}

\item {The time interval, $\Delta T$, between the Type II launch time and the flare onset.
As the emission starts some minutes after the onset of the flare impulsive phase,
back-extrapolation of the emission lanes in the ARTEMIS--IV spectra
was used to estimate the type II launch time.}
\end{itemize}

In order to quantify similarities between each pair (i,j) of events we computed,
a \textit{proximity measure} in the form of \textit{Standard Eucledean Distance\/}, $d_{ij}$; the smaller the 
index $d_{ij}$ between a pair the more its members resemble each other. We define $d_{ij}$ as:
\[
d_{ij}  = \sqrt {\left[ {{V_{II} } _i  - {V_{II} } _j } \right]^2  + \left[ {{V_{r}}_i  - {{V_{r}}_j } } \right]^2  + {\left[ {D_i  - D_j } \right]} ^2  + {\left[ {\delta \tau _i  - \delta \tau _j } \right]} ^2  + {\left[ {\Delta T_i  - \Delta T_j } \right]} ^2 }  
\]
Each coordinate (parameter) in the sum of squares is inversely weighted to the standard deviation of that coordinate.
The {\textit{proximity measure}} ($d_{ij}$)  between pairs of events is used as a criterion
for the identification of clusters within our data set. Certain groups of \textit{similar} events emerge thus
and are summarized in Table \ref{tab1}:

\begin{itemize}

\item{Group I (Events 25, 19 \& 14). Events in this group exhibit a close time
relationship between the type II launch and the flare onset (average $\Delta T$=-4.3$\pm$0.3 min);
the type IIs are fast (average $V_{II}$=1543$\pm$212 km/sec) while
the CME launch precedes the flare by 22$\pm$5.2 minutes and the CME speed is almost half the
shock speed (average $V_{r}$=$V_{II}/V_{CME}$=2.5$\pm$0.4). This suggests that
the type II radio source is located behind the leading edge of the CME and
that the associated shock was probably ignited by the flare and was propagating
through the transient disturbance at the wake of the CME. This is consistent with the
scenario proposed by \citet{WagnerMacQueen83} for the 17 April 1980 type II burst
and \citet{Vrsnak06} for the 3 November 2003 event (cf. also \citet{Vrsnak08} for a review).
Two of the three events  lack 
interplanetary type II in the WIND$/$WAVES reports, the only exception being Event 25.}

\item{Group II (Events 21, 30, 39, 32, 40, 27, 08, \& 23). Here the type II launch time is
well associated with the flare onset (average $\Delta T$=-0.90$\pm$0.20 min) and the CME launch which is
about 9.5$\pm$3.8 minutes after the flare. The type II speeds (average $V_{II}$=537$\pm$57 km/sec), on the other hand, are equal or less to the CME speeds (average $V_{r}$=$V_{II}/V_{CME}$=0.9$\pm$0.14); the type II duration
(average 5.8$\pm$1.3 min) more often than not exceeds the typical values for coronal type IIs.
It is expected that the type II are driven by CME bow shocks (when $V_{r}\approx1$) or CME flanks,
or are reconnection shocks induced by the CME liftoff. Three of the Events (39, 08 \& 23)
have an interplanetary type II as is often the case with shocks
driven by CME front or flanks. For the rest (21, 27, 30, 32 \& 40) no interplanetary type II was reported.}
\end{itemize}

The association between the events 33, 06 \& 36 remains at present uncertain, however they are characterized by fast CMEs
and long duration type II shocks.

\section{Discussion \& Conclusions}
We have studied fourteen complex events, each includes
a coronal type II burst a GOES SXR flare and a LASCO CME; certain
parameters, related to shock \& CME kinetics and radio bursts-flare-CME timing
were compared. Uncertainty factors were:

\begin{itemize}
\item{Projection Effects: They introduce inaccuracies in the CME speed calculation; the errors are
minimal in the case of limb CMEs.}
\item{Take off Time: Both CME onset and Type II start are estimated from backward extrapolation
neglecting possible acceleration.}
\item{Type II Speed: The calculations rely heavily on the coronal model adopted; this can be resolved
in the case of radio images of limb events.}
\item{SXR Flare Onset: Depends on the detection threshold used.}
\end{itemize}

Despite the uncertainties, it was found that most of the events may be grouped together based
on their similar behavior; for each group we have conjectured a
different shock generation processes, as various mechanisms exist (\cite{Vrsnak08}). 
Further study, with a larger data set of, preferably, limb events, 
will provide improved results as regards the coronal type II drivers.

\end{document}